\documentclass[a4paper,11pt]{article}
\usepackage{aaskaiid}
\usepackage{orcidlink}
\title{Time Domain Studies of Active Galactic Nuclei with the SKA telescopes}
\ShortTitle{Time Domain AGN}

\author[1,3]{Hayley Bignall\orcidlink{0000-0001-6247-3071}}
\ShortName{Hayley Bignall et al.} 
\author[2]{Rajan Chhetri\orcidlink{0000-0002-6966-9357}}
\author[]{the Transient Science Working Group}

\affiliation[1]{Manly Astrophysics, 15/41-42 East Esplanade, Manly, NSW 2095, Australia}
\emailAdd{hayley.bignall@manlyastrophysics.org}
\affiliation[2]{CSIRO Space and Astronomy, PO Box 1130, Bentley, WA 6102, Australia}
\affiliation[3]{Visiting Scientist, CSIRO Space and Astronomy, PO Box 1130, Bentley, WA 6102, Australia}

\abstract{Time domain studies of active galactic nuclei (AGN) at radio wavelengths probe physical processes near the central engine via intrinsic variability, in particular within the relativistic jets, as well as small-scale structures in the local Galactic interstellar medium (ISM) via scintillation and scattering effects. Recent discoveries reinforce the expectation that the high sensitivity, large field-of-view, and broadband frequency coverage of the SKA telescopes will help to revolutionise our understanding of AGN populations and the evolution of jets, and allow detailed modelling of the structure and dynamics of scattering plasma in the local ISM over a large fraction of the sky.} 

\begin{document}
\maketitle
\newcommand{\actaa}{Acta Astron.} 
\newcommand{\araa}{ARA\&A} 
\newcommand{\aar}{A\&ARv} 
\newcommand{\aapr}{A\&ARv} 
\newcommand{\ab}{Astrobiol.} 
\newcommand{\aj}{AJ} 
\newcommand{\apj}{ApJ} 
\newcommand{\apjl}{ApJL} 
\newcommand{\apjs}{ApJSS} 
\newcommand{\ao}{Appl. Opt.} 
\newcommand{\apss}{Astro. \& Space Sci.} 
\newcommand{\aap}{A\&A} 
\newcommand{\aaps}{A\&AS.} 
\newcommand{\baas}{Bull. Am. Astron. Soc.} 
\newcommand{\caa}{Chinese A\&A} 
\newcommand{\cjaa}{Chinese J. A\&A} 
\newcommand{\cqg}{Class. Quantum Gravity} 
\newcommand{\gal}{Galaxies} 
\newcommand{\gca}{Geo. Cosmo. Acta} 
\newcommand{\icarus}{Icarus} 
\newcommand{\jcap}{JCAP} 
\newcommand{\jgr}{J. Geophys. Res.} 
\newcommand{\jgrp}{J. Geophys. Res. Planets} 
\newcommand{\jqsrt}{J. Quant. Spectrosc. Radiat. Transf.} 
\newcommand{\memsai}{Mem. SAIt} 
\newcommand{\mnras}{MNRAS} 
\newcommand{\nat}{Nature} 
\newcommand{\nastro}{Nat. Astron.} 
\newcommand{\ncomms}{Nat. Commun.} 
\newcommand{\nphys}{Nat. Phys.} 
\newcommand{\na}{New Astron.} 
\newcommand{\nar}{New Astron. Rev.} 
\newcommand{\physrep}{Phys. Rep.} 
\newcommand{\pra}{Phys. Rev. A} 
\newcommand{\prb}{Phys. Rev. B} 
\newcommand{\prc}{Phys. Rev. C} 
\newcommand{\prd}{Phys. Rev. D} 
\newcommand{\pre}{Phys. Rev. E} 
\newcommand{\prx}{Phys. Rev. X} 
\newcommand{\prl}{Phys. Rev. Let.} 
\newcommand{\psj}{Planet. Sci. J.} 
\newcommand{\planss}{Planet. Space Sci.} 
\newcommand{\pnas}{Proc. Natl Acad. Sci. USA} 
\newcommand{\procspie}{Proc. SPIE} 
\newcommand{\pasa}{PASA} 
\newcommand{\pasj}{PASJ} 
\newcommand{\pasp}{PASP} 
\newcommand{\rmxaa}{RMXAA} 
\newcommand{\sci}{Science} 
\newcommand{\sciadv}{Sci. Adv.} 
\newcommand{\solphys}{Sol. Phys.} 
\newcommand{\sovast}{Soviet Ast.} 
\newcommand{\ssr}{Space Sci. Rev.} 
\newcommand{\uni}{Universe} 

\section{Introduction}
Radio studies of active galactic nuclei (AGN) in the time domain reveal a wealth of information about processes near the central supermassive black hole, as well as 
propagation effects in intervening media along the line-of-sight. 
\citet{2015aska.confE..58B} outlined the breadth of science that can be undertaken via time domain studies of AGN with SKA-Mid, which remains essentially valid. AGN dominate the radio sky, and indeed the dynamic radio sky \citep{2023MNRAS.523.2219A}, and the SKA telescopes will provide an immense legacy dataset for studies of variability across a broad range of timescales, frequencies, and object classes. 
Such a dataset would address multiple scientific questions, only a subset of which are mentioned in the present chapter. 
It is important to note that the interpretation of AGN variability at radio wavelengths requires a knowledge of both intrinsic and propagation effects. 
A number of relevant discoveries have been made in the past decade---some of which were quite unexpected---which further reveal the value of widefield, broadband and multi-timescale studies of AGN at radio wavelengths. Moreover, exploration of the time domain at other wavelengths is currently undergoing enormous expansion with the advent of telescopes such as the Vera C.\ Rubin Observatory in the optical  \citep[e.g.][]{2022ApJ...941...41C} and the Cherenkov Telescope Array Observatory for high energy gamma rays
\citep[e.g.][]{PassosReis:2025jii}; thus the realisation of the SKA telescopes is very timely with respect to multi-wavelength and multi-messenger studies of AGN. 
The present chapter discusses a few recent discoveries of relevance, and further elucidates what impact the SKA telescopes 
will have for understanding both AGN and intervening scattering media.

We attempt to avoid repetition of the discussion already presented in \citet{2015aska.confE..58B}, and here we also note several related topics which are covered elsewhere in the present volume.
\begin{itemize}
\item The study of radio flares/jet launching associated with Tidal Disruption Events \citep{TaoAn03.2026.SKA,Shu01.2026.SKA} is very important for understanding jet formation.
\item Multiwavelength and multi-messenger synergies \citep{Castignani01.2026.SKA, Rosch01.2026.SKA}, and high angular resolution studies with VLBI \citep{Kadler01.2026.SKA} are important to understand the high energy emission processes in relativistic jets.
\item In the present chapter we discuss interstellar scintillation. It is worth noting that measurements of interplanetary scintillation \citep{Chhetri01.2026.SKA} also provide information beyond the resolution of the SKA telescopes (without VLBI), on the population of sub-arcsecond scale AGN components.
\item Commensal image-plane search methods for detecting and analysing variable and transient sources are discussed by \citet{AlexAndersson01.2026.SKA}.
\item Studies of interstellar scattering of pulsars reveal much about the structure of the intervening media \citep{Tiburzi01.2026.SKA}, which is complementary to information from interstellar scintillation and scattering of AGN.
\end{itemize}

For a general overview of AGN jet physics,  \citet{2019ARA&A..57..467B} recently reviewed the current understanding of relativistic jets from AGN. \citet{2019NewAR..8701541H} reviewed multiwavelength observational properties of blazars, which are radio-loud AGN with a jet pointing close to the line-of-sight, including observations of variability, and how these can be used to constrain theoretical models. 
\citet{2019NatAs...3..387P} discussed the origin of radio emission from radio-quiet AGN, which represent $\sim 90$\% of the AGN population, and how the next generation radio telescopes, including the SKA, can probe radio emission from a wide range of possible mechanisms. Observations in the time domain provide important tests for the various mechanisms for radio emission. 
More recently, \citet{murphy2025dawesreview13new} present a broad overview of the dynamic radio sky, including AGN, and outlook toward future surveys for transient and variable radio sources.

Section 2 below briefly highlights various observational discoveries made in radio time domain studies of AGN in the past decade; in Section 3 we expand on some of the outstanding questions which the SKA telescopes will help to answer. In Section 4 we summarise the observational requirements for addressing the various questions, and present a brief summary and outlook in Section 5. 

\section{Some highlights from the past decade in time domain studies of AGN}
\label{sec:highlights}

The SKA pathfinder and precursor telescopes, along with broad-band and/or widefield upgrades to older telescopes, have enabled a number of exciting discoveries in time domain studies of AGN, giving us a taste of what will be possible with the greater sensitivity, survey speed and frequency coverage of the SKA telescopes. 

\subsection{Intra-hour variability near 1~GHz}
\label{sec:idv}
For more than 30 years since the discovery of radio intra-day variability (IDV) of AGN \citep{1987AJ.....94.1493H}, the largest amplitude IDV was typically observed at frequencies of a few GHz, with timescales becoming shorter toward higher frequencies \citep{1997ApJ...490L...9K,2001MNRAS.325.1411K}. This IDV has been shown to be predominantly due to interstellar scintillation (ISS) \citep{2008ApJ...689..108L}. 
Variability at $\sim 1$~GHz and below is typically slower (days to weeks or longer), due to refractive scintillation in the strong scattering regime. Previously studied, high flux density AGN are for the most part too large to exhibit diffractive scintillation, as the variations are averaged out over the angular diameter of the source \citep{1981ApJ...246...91D}, although \citet{2006A&A...446..185M} found evidence for a diffractive scintillation contribution to dynamic spectra of the fast scintillating quasar J1819+3845 around 1.4\,GHz (occurring due to a very nearby foreground scattering screen).

Hence, it was somewhat surprising when \citet{2020A&A...641L...4O} found an AGN showing $\sim 50$\% variations on a timescale of 6.5 minutes at 1.4~GHz with Apertif. In fact the source was discovered due to its variability producing prominent artefacts seen in the image integrated over 11 hours. 
At around the same time, \citet{2021MNRAS.502.3294W} made the remarkable discovery with ASKAP (operating with a 288~MHz bandwidth at a central frequency of 945~MHz) of 5 intra-hour variable AGN lying along a straight line, inferred to correspond to a narrow plasma filament within $\sim 10$\,pc of the Earth.  
In the cases above, follow-up observations revealed clear annual cycles in variability timescale, a result of the Earth's orbital motion with respect to the scattering screen. In all cases the scintillation pattern was determined to be highly anisotropic, which is also inferred from observations of pulsar secondary spectra \citep{10.1111/j.1365-2966.2004.08159.x}. The small inferred scattering screen distances relax the maximum angular size for a source to be sufficiently compact to exhibit scintillation, and give rise to rapid rates of scintillation, compared with more distant scattering material. 

The sensitivity of SKA-Mid will enable nearby scattering screens to be delineated over most of the sky with high accuracy, and their kinematics to be determined (Section~\ref{sec:skascreens}), providing clues to their origin.

\subsection{Identifying Extreme Scattering Events in broadband spectra}
Extreme Scattering Events (ESEs) were first detected as large flux density variations with a characteristic lens-like pattern, typically lasting several months in radio light curves of quasars. ESEs are interpreted as refractive lensing by AU-scale structures in the Galactic interstellar medium (ISM) \citep{1987Natur.326..675F}. ESEs are highly chromatic, often displaying caustic focusing features at centimetre wavelengths, and \citet{2016Sci...351..354B} exploited the wideband CABB backend at the Australia Telescope Compact Array to identify ESEs via features in instantaneous spectra, and subsequently follow them up with dense monitoring every few days to obtain a dynamic spectrum (the ATESE project). ESEs are relatively rare, with roughly one event in a few hundred highly compact sources expected in progress at any given time; the project of Bannister et al.\  monitored approximately 1000 sources and detected a handful of ESEs over its lifetime. A novel technique was developed for using the broadband dynamic spectra over the course of the ESE to map the column density profile of the plasma lens \citep{2016ApJ...817..176T}. 
The ATESE project also led to the discovery of unusual spectral features in a scintillating AGN that were not ascribable to an ESE, but rather to a possible combination of refractive and diffractive effects  \citep{2017MNRAS.469.5023T}.
As the origin of the ESE lenses is not yet known, nor even the event rate accurately determined, measurement of a larger number of events through monitoring many thousands of compact AGN would greatly advance this field (Section~\ref{sec:skascreens}).

\subsection{Symmetric Achromatic Variations}
\citet{2017ApJ...845...89V} identified symmetric U-shaped features in light curves on year-long timescales, initially at 15\,GHz from the long term OVRO monitoring program. These variations are reminiscent of ESEs but were found to be achromatic over a large range of frequencies, which is not consistent with plasma lensing. The authors suggest that gravitational lensing, wherein compact source components cross the magnification pattern of a gravitational lens at relativistic speeds, may be responsible for the observed variations. 
Monitoring a much larger number of AGN with SKA-Mid would turn up many more examples of this rare phenomenon, allowing models for its origin to be tested. 
 
\subsection{Variability at SKA-Low frequencies}
\label{sec:kat}
Using two epochs of observations from the GaLactic and Extragalactic All-sky Murchison Widefield Array (GLEAM) survey that were taken one year apart, \citet{2021MNRAS.501.6139R} searched for spectral variability across 100-230 MHz for 21,558 sources, finding 323 sources with significant spectral variability. Subsequently, 15 low frequency variable peaked-spectrum sources were studied in a handful of epochs spread over several months, across the range 0.072-10\,GHz \citep{2022MNRAS.512.5358R}. No variability was detected at 1-10\,GHz, while the majority of sources exhibited significant variability at lower frequencies, indicating that the low frequency variability is predominantly due to refractive ISS of source components $\sim 25$\,mas in extent. The authors argue that broadband spectral variability can be used to determine the origin of variability; distinguishing among scintillation, variable absorption, and variations in the jet. 
The broad bandwidth of the SKA telescopes will thus be useful to determine the likely origin of observed variability for candidate AGN, even when not well sampled in the time domain. At SKA-Low frequencies, and the lower bands of SKA-Mid, refractive interstellar scintillation is expected to be the predominant cause of time domain variability in the measured flux density of AGN.

\subsection{Compact Symmetric Objects}
\label{sec:csos}
Compact Symmetric Objects (CSOs) are extragalactic objects with regions of emission on both sides of an active galactic nucleus, are $<1$\,kpc in total extent, and are not dominated by relativistic beaming effects. They are characterised by their low variability over timescales of years \citep{Kiehlmann_2024}. Although CSOs have been known for decades, it has been challenging to identify well-defined samples of CSOs, in part due to the dominance of beamed (blazar-type) objects in flux density limited surveys of compact sources; yet CSOs are likely very important for understanding AGN central engines and radio jet formation \citep{Readhead2024}.
\citet{2024ApJ...961..241K} demonstrated that the vast majority of CSO 2s (edge-brightened CSOs) are short-lived and do not evolve into larger radio galaxies. Only relatively bright CSOs have been confirmed and studied in detail so far, although low-power samples are being investigated \citep{2025A&A...698A.157O}. Using their low variability characteristics together with spectral information (peaked-spectrum) may be an effective way to identify candidate CSOs from compact (on arcsecond-scale) source samples. In the absence of very high angular resolution (VLBI) observations, interplanetary scintillation \citep{Chhetri01.2026.SKA} and low-frequency refractive ISS (Section~\ref{sec:kat}) may provide efficient means of identifying a large number of CSO candidates from SKA surveys. 


\subsection{Extreme radio variability in Narrow-Line Seyfert 1 galaxies}
\label{sec:nlsy1}
\citet{2024MNRAS.532.3069J} report on extreme variability observed at 37\,GHz, as large as 3-4 orders of magnitude on timescales of days, in normally radio-weak narrow-line Seyfert 1 (NLS1) galaxies. The authors consider a variety of possible mechanisms to explain the variability, determining the most plausible possibilities to be (1) flaring due to a shock generated by the interaction between a jet and cloud or star, (2) variable absorption of a jet within the broad-line region, or (3) magnetic reconnection in the jet or black hole magnetosphere. Observations across SKA-Mid band 5, and ideally also simultaneously in other parts of the electromagnetic spectrum, would help to determine the mechanism responsible for the observed extreme radio variability. 


\section{Outstanding questions for time domain studies of AGN}
The discoveries mentioned in Section~\ref{sec:highlights} highlight the value of high sensitivity, broad frequency coverage and wide field of view for variability studies. In this section we discuss how AGN variability observed with the SKA telescopes can address a number of outstanding questions.

\subsection{What are the interstellar scattering screens?}
ESEs and ISS indicate the presence of small, discrete structures in the ISM. There is still no conclusive explanation for all of the observed phenomena, but theoretical work in this field is ongoing; a couple of recent contributions are mentioned below. It is worth noting that radio observations are the only known method to probe these otherwise unseen, small-scale structures.

\citet{2024MNRAS.528.6292J}, and references therein, outline the current problems for explaining ESEs. The authors show how ESE features arise when the lens has a cusp-like profile, extending previous work to show that cusp as well as fold catastrophes can be produced by thin corrugated plasma sheets viewed in projection. They propose that this framework could provide a universal model for both ESEs and scintillation. 

The long, straight, narrow plasma filament implied by the discovery of a line of fast scintillating AGN by  \citet{2021MNRAS.502.3294W} (Section~\ref{sec:idv}) is reminiscent of tidal stream debris, and   
prompted theoretical work by \citet{2025ApJ...982...61S}
to model tidal disruption of molecular hydrogen ``snow clouds'' by unassociated stars. The authors showed that microstructure in the resulting tidal streams has properties consistent with those inferred from ISS. 
If the clouds are unassociated with the disrupting star, the expected event rate is much too low to explain the prevalence of ISS in general, however this may be resolved if most snow clouds are loosely gravitationally bound to host stars.

\subsubsection{How SKA-Mid will help}
\label{sec:skascreens}
Regarding expectations for the prevalence of ISS and ESEs among lower flux density ($\lesssim 1$\,mJy) sources, it is important to note that: \\
(i) as well as increasing source density with lower limiting flux density, fainter sources can be more compact for a given brightness temperature than their bright counterparts, hence ISS and ESEs are more likely to be observed, provided the flux density is dominated by a compact component. For example, a 1\,mJy source with intrinsic brightness temperature $10^{11}$\,K at 1\,GHz would have angular size $\sim 0.1$\,mas (in the absence of scatter-broadening) and would be likely to scintillate through a scattering ``screen'' within 100\,pc, being only a few times larger than the Fresnel scale at that distance. Most lines of sight could be expected to intersect significant scattering material within 100\,pc, based on the prevalence of scattering screens inferred from pulsar observations \citep{2025NatAs...9.1053R}.\\
(ii) significant numbers of rapidly scintillating AGN are being found behind very nearby (within $\sim 10$\,pc) scattering screens which seem to be clustered over several degrees on the sky \citep{2023MNRAS.523.5661W}. Higher sensitivity provides a higher source density to map out these nearby screens and determine their structure and kinematics, which is expected to provide clues as to their origin. 

A wide field of view allows multiple scintillating sources to be observed simultaneously, greatly increasing observing efficiency.
Although screens within 10\,pc only cover a small fraction of the sky, it is anticipated that thousands of rapidly scintillating sources will be found behind such screens with SKA-Mid.
ASKAP is able to detect significant intra-hour variability for source flux densities down to $\sim 1$\,mJy; SKA-Mid AA4 will go more than an order of magnitude fainter, albeit with a smaller field of view. An observational strategy to cover a larger sky area efficiently would be to cycle frequently among several pointings, sacrificing continuous light curves; variability characteristics can  still be measured with reasonable accuracy. 
Observations at several epochs over the course of a year are needed to measure annual cycles in the characteristic rate of ISS; this determines the geometry (angle and degree anisotropy in the scintillation pattern), and velocity of the scattering plasma, and constrains the distance to the scattering screen  \citep{2021MNRAS.502.3294W}. Annual cycles for multiple sources allow the investigation of the scale and coherence of associated, albeit patchy, scattering structures, in particular locations in the nearby ISM. 

ESEs are distinguished from ISS in their characteristics and require different observing strategies, but it seems likely the small-scale ISM structures responsible for both phenomena are related. Both phenomena are highly frequency dependent and modelling them accurately requires observations over a large frequency range.
The deployment of band 3 and 4 receivers for SKA-Mid would be advantageous for the modelling of ESEs in particular, to close the gap in frequency coverage between 1.76 and 4.6\,GHz, as ESEs typically have strong structure in this range. In the absence of these bands, any or all of bands 2, 5a and 5b could be used to identify ESEs by searching for caustic features in the broadband spectra, ideally in near real-time to enable follow-up of candidate events in progress, and/or analysing light curves from regular, relatively shallow monitoring observations of large sky areas. Once identified, follow-up observations at high angular resolution and at other wavelengths could be initiated to confirm and study candidate ESEs in detail, including refractive shifts and/or multiple imaging from VLBI which are needed to determine lens geometries \citep{2016Sci...351..354B}.

It is also worth noting that broad-band observations of ISS, across the range of SKA-Mid, will be extremely useful to determine the strength of scattering via the frequency dependence, enabling unambiguous modelling of the nearby scattering screen properties toward each scintillating source (this was not possible from the ASKAP data alone published by \citet{2021MNRAS.502.3294W}). Annual cycles can be measured using a single band, while single-epoch observations in multiple bands, observed close together in time during a rapid scintillation phase, would suffice to determine the frequency dependence.

In summary, with SKA-Mid we will be able to determine ESE event rates and lens characteristics, and trace out regions of enhanced (or reduced) scattering in the ISM, along with their geometry, microstructure and dynamics, with far greater accuracy than ever before. This will enable detailed comparisons with expectations of the various models, and help to determine the physical origin of the scattering screens. We note that at low Galactic latitudes, both ESEs and rapid ISS are likely to be suppressed on many lines of sight due to scatter-broadening in more distributed plasma \citep{2022MNRAS.515.1736K}.

\subsection{What is the compact AGN contribution to the sub-mJy radio source population?}
To date, there is no evidence for a reduction in the fraction of very compact components that can exhibit variability or scintillation, down to the mJy source population \citep{2014AJ....147...14D}, although around 0.1\,mJy, jetted AGN make up a much smaller fraction of the radio source counts \citep{2016A&ARv..24...13P}. 
Statistics of scintillation as a function of flux density for large samples would provide constraints on the fraction of highly compact radio sources. 

\subsubsection{How the SKA telescopes will help}
The SKA telescopes will detect many scintillating AGN, both blazars and unbeamed sources---potentially including faint CSOs and newly formed radio jets. 
AGN found to be scintillating will be among the most compact. Thus, studies of ISS with the SKA telescopes will probe parsec- and sub-parsec scale properties of hundreds of thousands of compact sources, without requiring very long baseline observations and at even higher resolution than provided by ground-based VLBI.

SKA-Low and SKA-Mid Band 1 would predominantly see refractive scintillation on timescales of months in sources dominated by milliarcsecond-scale components.  Thus, infrequent monitoring, e.g. several epochs over the course of a year or more, of the observable sky would provide spectral and flux density variations for millions of sources, allowing an estimate of compact fraction on milliarcsecond scales. 
Statistics of ISS at higher frequencies go to even smaller scales (10-100\,$\mu$as). 
As discussed in \citet{Chhetri01.2026.SKA}, interplanetary scintillation (IPS) can be used to determine the compact fraction, albeit with a higher flux density limit, on sub-arcsecond scales, requiring only short observations because of the sub-second flux variations induced by IPS.  

\subsection{What are the mechanisms for intrinsic variability in various classes of AGN?}
Most long-term variability studies to date have been of relatively bright blazars \citep[e.g.][]{2022MNRAS.510..469K,2024A&A...681A..69B,2025A&A...693A.318K}, and a few nearby radio-quiet quasars have also been studied in detail \citep[e.g.][]{2020ApJ...891...59R}. This has limited our understanding of the range of AGN variability, lifetimes and processes occurring in relativistic jets.

\subsubsection{How the SKA telescopes will help}
The SKA telescopes will probe variability down to much lower flux densities, in different populations of AGN. Only very compact sources exhibit variability on human timescales. Addressing what drives the radio emission processes in AGN is of great importance to understanding not only the physical processes within jets, but their role in galaxy evolution via feedback mechanisms.  
Statistics of variability (or lack of) from low to extreme levels can probe compactness and relativistic beaming. A low level of variability, for example, can be used as a criterion to identify candidate CSOs \citep{Kiehlmann_2024}.

Observations at the high frequency end of SKA-Mid, in Band 5, would be most likely to detect intrinsic events such as those mentioned in Section~\ref{sec:nlsy1}; indeed most intrinsic AGN variability is expected to be more extreme at higher radio frequencies. Toward lower frequencies,  below $\sim 10$\,GHz, flares are often dampened by absorption effects (synchrotron self-absorption, or free-free absorption), meaning they tend to be slower, peaking at later times and with reduced amplitude. 
Identifying flaring behaviour in near real-time could be used to trigger VLBI observations including SKA-Mid, which would allow discrimination among various models for the ejected plasma \citep[e.g.][]{2020ApJ...891...59R}.
 The origin of rare phenomena such as Symmetric Achromatic Variations could also be investigated; finding more examples of such variability is crucial. 
 SKA-Mid could monitor potentially millions of AGN. There would be trade-offs for total telescope time required for large area, shallow surveys, vs deeper surveys requiring longer integration times, as well as observing cadence. 

\section{Observational and data requirements}
Below we summarise requirements and desired data products from the SKA telescopes for the various applications discussed in the present chapter.

\subsection{Long-term variations (months to years)}
Something like the Rapid ASKAP Continuum Survey \citep{McConnell_Hale_Lenc_Banfield_Heald_Hotan_Leung_Moss_Murphy_O’Brien_etal._2020}, i.e.\ a relatively fast and shallow all-sky survey, repeated on a quasi-regular basis would be extremely valuable for investigating AGN intrinsic variability (preferably SKA-Mid Band 5) and slow refractive scintillation (SKA-Low and SKA-Mid Band 1), among many other transient science applications, across the entire observable sky. 
Deeper observations with narrower sky coverage would also be valuable to explore the variability statistics and compactness of the faint ($\sim 10\mu$Jy) source population. 
Integrated images with wide frequency channels would be sufficient to generate source catalogues for such variablity studies not requiring particularly high time or frequency resolution. 

\subsection{Short timescale variations (minutes to hours)}
To find and monitor rapid ISS, it would be desirable to produce light curves with a time resolution of order a minute from longer observations. The power spectra of fluctuations from ISS encode information on source structure on microarcsecond scales, smaller than resolvable even with VLBI \citep{2007MNRAS.380L..20M}. 
Measuring annual cycles in scintillation rate for rapidly scintillating sources, to determine the dynamics and anisotropy of scattering with reasonable accuracy,  requires observations of the target fields over several hours at $\approx 6$ epochs over the course of a year \citep[e.g.][section \ref{sec:skascreens}]{2021MNRAS.502.3294W}. 

\subsection{Variations on intermediate timescales (days to weeks)}
Variability on timescales of days to weeks at GHz frequencies is expected to be dominated by refractive interstellar scintillation \citep[e.g][]{2008ApJ...689..108L}; ESEs also show changes on these timescales. 
ESE candidates may be identified either through light curve behaviour and/or spectra showing large deviations from a power law. Ideally, ESEs found to be in progress would be monitored with a near-daily cadence over a large range of frequencies, to model the plasma density variations from the dynamic spectrum \citep{2016ApJ...817..176T} (see Section~\ref{sec:dedicated}). 

\subsection{Frequency coverage and resolution}
Observations at a wide range of frequencies are crucial to discriminate among various models for essentially all of the areas discussed above in Section~\ref{sec:highlights}.

Modest frequency resolution ($\sim 1$\% of the sky frequency) is desirable to study diffractive scintillation and ESE caustics, which are highly frequency dependent. 

For detailed modelling of ESEs and ISS, SKA-Mid Bands 3 and 4 would be highly desirable to close the gap in frequency coverage between 1.76 and 4.6~GHz, as the most extreme variability of AGN due to interstellar propagation is often seen in this region \citep{2001MNRAS.325.1411K}. 

A future upgrade to extend SKA-Mid to 25\,GHz, where variations are less suppressed due to optical depth effects, would be advantageous for studies of AGN intrinsic variability and connections with multiwavelength/multi-messenger observations including high energy gamma-ray and neutrino events. 

\subsection{Polarization}
\citet{2019NewAR..8701541H} emphasise the value of full polarization spectra, including sensitive circular polarization observations, for understanding particle acceleration processes and particle composition of jets. Polarized components of AGN can be more compact than the total intensity components, meaning they may also show large and rapid scintillations \citep{2000ApJ...538..623M,2002ApJ...581..103R}.

\subsection{High angular resolution with VLBI}
ESEs are expected to produce refractive shifts, image distortion and/or multiple imaging which can be studied with high angular resolution VLBI observations including SKA-Mid, and potentially SKA-Low. Such observations are extremely valuable to discriminate among ESE lens models \citep{2016Sci...351..354B}. 
High angular resolution VLBI observations, particularly at the higher Band 5 frequencies, are also important to understand AGN jet variability in detail \citep[e.g.][]{2020ApJ...891...59R}.

\subsection{Dedicated follow-up/monitoring, including with other facilities} \label{sec:dedicated}
As discussed in \citet{2015aska.confE..58B}, sub-arraying of the SKA telescopes could be used both to increase frequency coverage and/or sky area at the cost of reduced sensitivity, or potentially 
allow dedicated monitoring of selected phenomena of particular interest with a small subset of SKA antennas/stations.

Although the SKA telescopes, particularly SKA-Mid, will be hugely powerful for AGN variability studies, delving into flux density regimes and source types never before investigated dynamically, 
there remains an important role for less sensitive arrays and single dishes that can perform more dedicated, targeted monitoring, albeit of a relatively small selection of brighter sources/events and/or at select frequencies. 
Subarraying SKA-Mid, or using a different monitoring facility \citep{fender2024fillingradiotransientsgap}, e.g.\ for sufficiently bright ESEs, may be a practical means of obtaining the desired dynamic spectra for these events. 

The decades-long, dedicated monitoring programs undertaken at facilities such as the University of Michigan Radio Observatory, Mets\"{a}hovi Radio Observatory, and Owens Valley Radio Observatory are still proving extremely valuable; for example, \citet{2026ApJ...996L..39R} report the discovery of a fundamental and a harmonic periodicity in the light curve of blazar PKS~J1309+1154, making it a strong supermassive black hole binary candidate likely to be a source of gravitational waves. \citet{2025A&A...693A.319K} point out that very long monitoring programs (decades) may be needed to characterise radio variability for many AGN; this should be taken into account when interpreting flux density state and variability from limited time sampling. 

\subsection{Effect of staged delivery of the SKA telescopes}
Given the huge number of observable AGN, valuable time domain studies can be undertaken with the AA* realisation of the SKA telescopes. When the array is enhanced, the fainter source population will be probed.

\section{Summary}
We have outlined how observations of the dynamic radio sky with SKA telescopes will address some of the outstanding problems across a broad range of topics, from AGN physics to small-scale phenomena in the Galactic ISM. 
A large variety of phenomena are observed in time domain studies of AGN, and there is much remaining to be understood, both for intrinsic mechanisms of variability as well as variability due to propagation effects. AGN intrinsic variability is key to understanding the conditions under which relativistic jets form, expand and interact with the surrounding medium in their host galaxies; while variability due to interstellar scattering reveals ubiquitous, mysterious small-scale structure which is not yet understood, but which may represent a significant fraction of baryons in the Galaxy \citep[][and references therein]{2025ApJ...982...61S}.

\bibliographystyle{abbrvnat-maxbibnames4}
\bibliography{chapter}

@ARTICLE{2020ApJ...891...59R,
       author = {{Reynolds}, Cormac and {Punsly}, Brian and {Miniutti}, Giovanni and {O'Dea}, Christopher P. and {Hurley-Walker}, Natasha},
        title = "{Estimating the Jet Power of Mrk 231 during the 2017-2018 Flare}",
      journal = {\apj},
         year = 2020,
        month = mar,
       volume = {891},
       number = {1},
          eid = {59},
        pages = {59},
          doi = {10.3847/1538-4357/ab72f0},
archivePrefix = {arXiv},
       eprint = {2001.10697},
 primaryClass = {astro-ph.GA},
       adsurl = {https://ui.adsabs.harvard.edu/abs/2020ApJ...891...59R}
}

@ARTICLE{2024A&A...681A..69B,
       author = {{Benke}, P. and {R{\"o}sch}, F. and {Ros}, E. and {Kadler}, M. and {Ojha}, R. and {Edwards}, P.~G. and {Horiuchi}, S. and {Hyland}, L.~J. and {Phillips}, C. and {Quick}, J.~F.~H. and {Stevens}, J. and {Tzioumis}, A.~K. and {Weston}, S.},
        title = "{TANAMI: Tracking active galactic nuclei with austral milliarcsecond interferometry. III. First-epoch S band images}",
      journal = {\aap},
         year = 2024,
        month = jan,
       volume = {681},
          eid = {A69},
        pages = {A69},
          doi = {10.1051/0004-6361/202347823},
archivePrefix = {arXiv},
       eprint = {2310.10206},
 primaryClass = {astro-ph.HE},
       adsurl = {https://ui.adsabs.harvard.edu/abs/2024A&A...681A..69B}
}

@ARTICLE{2026ApJ...996L..39R,
       author = {{Readhead}, A.~C.~S. and {Aller}, M.~F. and {Sullivan}, A.~G. and {Blandford}, R.~D. and {Mr{\'o}z}, P. and {De la Parra}, P.~V. and {Molina}, B. and {Most}, E.~R. and {Lister}, M.~L. and {Synani}, A. and et al.},
        title = "{Compelling Evidence for a Harmonic in the Light Curve of the Supermassive Black Hole Binary Candidate PKS J1309+1154}",
      journal = {\apjl},
         year = 2026,
        month = jan,
       volume = {996},
       number = {2},
          eid = {L39},
        pages = {L39},
          doi = {10.3847/2041-8213/ae2656},
archivePrefix = {arXiv},
       eprint = {2511.09409},
 primaryClass = {astro-ph.HE},
       adsurl = {https://ui.adsabs.harvard.edu/abs/2026ApJ...996L..39R}
}

@ARTICLE{2025A&A...693A.318K,
       author = {{Kankkunen}, S. and {Tornikoski}, M. and {Hovatta}, T. and {L{\"a}hteenm{\"a}ki}, A.},
        title = "{Long-term radio variability of active galactic nuclei at 37 GHz}",
      journal = {\aap},
         year = 2025,
        month = jan,
       volume = {693},
          eid = {A318},
        pages = {A318},
          doi = {10.1051/0004-6361/202450561},
archivePrefix = {arXiv},
       eprint = {2412.08191},
 primaryClass = {astro-ph.GA},
       adsurl = {https://ui.adsabs.harvard.edu/abs/2025A&A...693A.318K}
}

@ARTICLE{2025A&A...693A.319K,
       author = {{Kankkunen}, S. and {Tornikoski}, M. and {Hovatta}, T.},
        title = "{Active galactic nucleus time-variability analysis and its caveats}",
      journal = {\aap},
         year = 2025,
        month = jan,
       volume = {693},
          eid = {A319},
        pages = {A319},
          doi = {10.1051/0004-6361/202450562},
archivePrefix = {arXiv},
       eprint = {2412.08192},
 primaryClass = {astro-ph.GA},
       adsurl = {https://ui.adsabs.harvard.edu/abs/2025A&A...693A.319K}
}

@ARTICLE{2000ApJ...538..623M,
       author = {{Macquart}, J.-P. and {Kedziora-Chudczer}, Lucyna and {Rayner}, David P. and {Jauncey}, David L.},
        title = "{Strong, Variable Circular Polarization in PKS 1519-273}",
      journal = {\apj},
         year = 2000,
        month = aug,
       volume = {538},
       number = {2},
        pages = {623-627},
          doi = {10.1086/309184},
       adsurl = {https://ui.adsabs.harvard.edu/abs/2000ApJ...538..623M}
}

@article{McConnell_Hale_Lenc_Banfield_Heald_Hotan_Leung_Moss_Murphy_O’Brien_etal._2020,
    title={The Rapid ASKAP Continuum Survey I: Design and first results}, 
    volume={37}, 
    DOI={10.1017/pasa.2020.41},
    journal={Publications of the Astronomical Society of Australia},
    author={McConnell, D. and Hale, C. L. and Lenc, E. and Banfield, J. K. and Heald, George and Hotan, A. W. and Leung, James K. and Moss, Vanessa A. and Murphy, Tara and O’Brien, Andrew and et al.}, 
    year={2020}, 
    pages={e048}}

@misc{fender2024fillingradiotransientsgap,
      title={Filling the radio transients gap (or: The case for a dedicated radio transients monitoring array in the southern hemisphere)}, 
      author={Rob Fender and Assaf Horesh and Phil Charles and Patrick Woudt and James Miller-Jones and Joe Bright},
      year={2024},
      eprint={2402.04698},
      archivePrefix={arXiv},
      primaryClass={astro-ph.IM},
      url={https://arxiv.org/abs/2402.04698}, 
}

@ARTICLE{2002ApJ...581..103R,
       author = {{Rickett}, Barney J. and {Kedziora-Chudczer}, Lucyna and {Jauncey}, David L.},
        title = "{Interstellar Scintillation of the Polarized Flux Density in Quasar PKS 0405-385}",
      journal = {\apj},
         year = 2002,
        month = dec,
       volume = {581},
       number = {1},
        pages = {103-126},
          doi = {10.1086/344167},
archivePrefix = {arXiv},
       eprint = {astro-ph/0208307},
 primaryClass = {astro-ph},
       adsurl = {https://ui.adsabs.harvard.edu/abs/2002ApJ...581..103R}
}

@ARTICLE{2022MNRAS.510..469K,
       author = {{Kramarenko}, I.~G. and {Pushkarev}, A.~B. and {Kovalev}, Y.~Y. and {Lister}, M.~L. and {Hovatta}, T. and {Savolainen}, T.},
        title = "{A decade of joint MOJAVE-Fermi AGN monitoring: localization of the gamma-ray emission region}",
      journal = {\mnras},
         year = 2022,
        month = feb,
       volume = {510},
       number = {1},
        pages = {469-480},
          doi = {10.1093/mnras/stab3358},
archivePrefix = {arXiv},
       eprint = {2106.08416},
 primaryClass = {astro-ph.HE},
       adsurl = {https://ui.adsabs.harvard.edu/abs/2022MNRAS.510..469K}
}

@ARTICLE{2022MNRAS.515.1736K,
       author = {{Koryukova}, T.~A. and {Pushkarev}, A.~B. and {Plavin}, A.~V. and {Kovalev}, Y.~Y.},
        title = "{Tracing Milky Way scattering by compact extragalactic radio sources}",
      journal = {\mnras},
         year = 2022,
        month = sep,
       volume = {515},
       number = {2},
        pages = {1736-1750},
          doi = {10.1093/mnras/stac1898},
archivePrefix = {arXiv},
       eprint = {2201.04359},
 primaryClass = {astro-ph.GA},
       adsurl = {https://ui.adsabs.harvard.edu/abs/2022MNRAS.515.1736K}
}

@ARTICLE{2025NatAs...9.1053R,
       author = {{Reardon}, Daniel J. and {Main}, Robert and {Ocker}, Stella Koch and {Shannon}, Ryan M. and {Bailes}, Matthew and {Camilo}, Fernando and {Geyer}, Marisa and {Jameson}, Andrew and {Kramer}, Michael and {Parthasarathy}, Aditya and et al.},
        title = "{Bow shock and Local Bubble plasma unveiled by the scintillating millisecond pulsar J0437{\ensuremath{-}}4715}",
      journal = {Nature Astronomy},
         year = 2025,
        month = jul,
       volume = {9},
        pages = {1053-1063},
          doi = {10.1038/s41550-025-02534-6},
archivePrefix = {arXiv},
       eprint = {2410.21390},
 primaryClass = {astro-ph.HE},
       adsurl = {https://ui.adsabs.harvard.edu/abs/2025NatAs...9.1053R}
}

@ARTICLE{2007MNRAS.380L..20M,
       author = {{Macquart}, J.-P. and {de Bruyn}, A.~G.},
        title = "{Emergence and disappearance of microarcsecond structure in the scintillating quasar J1819+3845}",
      journal = {\mnras},
         year = 2007,
        month = sep,
       volume = {380},
       number = {1},
        pages = {L20-L24},
          doi = {10.1111/j.1745-3933.2007.00341.x},
archivePrefix = {arXiv},
       eprint = {0705.3414},
 primaryClass = {astro-ph},
       adsurl = {https://ui.adsabs.harvard.edu/abs/2007MNRAS.380L..20M}
}

@ARTICLE{2024ApJ...961..241K,
       author = {{Kiehlmann}, S. and {Readhead}, A.~C.~S. and {O'Neill}, S. and {Wilkinson}, P.~N. and {Lister}, M.~L. and {Liodakis}, I. and {Bruzewski}, S. and {Pavlidou}, V. and {Pearson}, T.~J. and {Sheldahl}, E. and {Siemiginowska}, A. and {Tassis}, K. and {Taylor}, G.~B.},
        title = "{Compact Symmetric Objects. II. Confirmation of a Distinct Population of High-luminosity Jetted Active Galaxies}",
      journal = {\apj},
         year = {2024b},
        month = feb,
       volume = {961},
       number = {2},
          eid = {241},
        pages = {241},
          doi = {10.3847/1538-4357/ad0cc2},
archivePrefix = {arXiv},
       eprint = {2303.11359},
 primaryClass = {astro-ph.HE},
       adsurl = {https://ui.adsabs.harvard.edu/abs/2024ApJ...961..241K}
}

@ARTICLE{2022ApJ...941...41C,
       author = {{Creque-Sarbinowski}, Cyril and {Kamionkowski}, Marc and {Zhou}, Bei},
        title = "{Active Galactic Nucleus Variability in the Age of Rubin}",
      journal = {\apj},
         year = 2022,
        month = dec,
       volume = {941},
       number = {1},
          eid = {41},
        pages = {41},
          doi = {10.3847/1538-4357/ac9eb2},
archivePrefix = {arXiv},
       eprint = {2110.13149},
 primaryClass = {astro-ph.GA},
       adsurl = {https://ui.adsabs.harvard.edu/abs/2022ApJ...941...41C}
}

@INPROCEEDINGS{2015aska.confE..58B,
       author = {{Bignall}, H.~E. and {Croft}, S. and {Hovatta}, T. and {Koay}, J.~Y. and {Lazio}, J. and {Macquart}, J.~P. and {Reynolds}, C.},
        title = "{Time domain studies of Active Galactic Nuclei with the Square Kilometre Array}",
    booktitle = {Advancing Astrophysics with the Square Kilometre Array (AASKA14)},
         year = 2015,
        month = apr,
          eid = {58},
        pages = {58},
          doi = {10.22323/1.215.0058},
archivePrefix = {arXiv},
       eprint = {1501.04627},
 primaryClass = {astro-ph.HE},
       adsurl = {https://ui.adsabs.harvard.edu/abs/2015aska.confE..58B}
}

@article{PassosReis:2025jii,
    author = "Passos Reis, Luana and others",
    title = "{Detection Prospects for AGNs with the Cherenkov Telescope Array}",
    doi = "10.22323/1.501.0798",
    journal = "PoS",
    volume = "ICRC2025",
    pages = "798",
    year = "2025"
}

@ARTICLE{2024MNRAS.532.3069J,
       author = {{J{\"a}rvel{\"a}}, E. and {Savolainen}, T. and {Berton}, M. and {L{\"a}hteenm{\"a}ki}, A. and {Kiehlmann}, S. and {Hovatta}, T. and {Varglund}, I. and {Readhead}, A.~C.~S. and {Tornikoski}, M. and {Max-Moerbeck}, W. and {Reeves}, R.~A. and {Suutarinen}, S.},
        title = "{Unprecedented extreme high-frequency radio variability in early-stage active galactic nuclei}",
      journal = {\mnras},
         year = 2024,
        month = aug,
       volume = {532},
       number = {3},
        pages = {3069-3101},
          doi = {10.1093/mnras/stae1701},
archivePrefix = {arXiv},
       eprint = {2312.02326},
 primaryClass = {astro-ph.GA},
       adsurl = {https://ui.adsabs.harvard.edu/abs/2024MNRAS.532.3069J}
}

@Article{Readhead2024,
	author        = {{Readhead}, A.~C.~S. and {Ravi}, V. and {Blandford}, R.~D. and {Sullivan}, A.~G. and {Somalwar}, J. and {Begelman}, M.~C. and {Birkinshaw}, M. and {Liodakis}, I. and {Lister}, M.~L. and {Pearson}, T.~J. and {Taylor}, G.~B. and {Wilkinson}, P.~N. and {Globus}, N. and {Kiehlmann}, S. and {Lawrence}, C.~R. and {Murphy}, D. and {O'Neill}, S. and {Pavlidou}, V. and {Sheldahl}, E. and {Siemiginowska}, A. and {Tassis}, K.},
	title         = {{Compact Symmetric Objects. III. Evolution of the High-luminosity Branch and a Possible Connection with Tidal Disruption Events}},
	journal       = {\apj},
	year          = {2024},
	volume        = {961},
	number        = {2},
	pages         = {242},
	month         = feb,
	adsurl        = {https://ui.adsabs.harvard.edu/abs/2024ApJ...961..242R},
	archiveprefix = {arXiv},
	doi           = {10.3847/1538-4357/ad0c55},
	eid           = {242},
	eprint        = {2303.11361},
	primaryclass  = {astro-ph.HE},
}

@ARTICLE{2019NewAR..8701541H,
       author = {{Hovatta}, Talvikki and {Lindfors}, Elina},
        title = "{Relativistic Jets of Blazars}",
      journal = {\nar},
         year = 2019,
        month = dec,
       volume = {87},
          eid = {101541},
        pages = {101541},
          doi = {10.1016/j.newar.2020.101541},
archivePrefix = {arXiv},
       eprint = {2003.06322},
 primaryClass = {astro-ph.HE},
       adsurl = {https://ui.adsabs.harvard.edu/abs/2019NewAR..8701541H}
}

@ARTICLE{2022MNRAS.512.5358R,
       author = {{Ross}, K. and {Hurley-Walker}, N. and {Seymour}, N. and {Callingham}, J.~R. and {Galvin}, T.~J. and {Johnston-Hollitt}, M.},
        title = "{Wide-band spectral variability of peaked spectrum sources}",
      journal = {\mnras},
         year = 2022,
        month = jun,
       volume = {512},
       number = {4},
        pages = {5358-5373},
          doi = {10.1093/mnras/stac819},
archivePrefix = {arXiv},
       eprint = {2203.11466},
 primaryClass = {astro-ph.GA},
       adsurl = {https://ui.adsabs.harvard.edu/abs/2022MNRAS.512.5358R}
}

@ARTICLE{2021MNRAS.502.3294W,
       author = {{Wang}, Yuanming and {Tuntsov}, Artem and {Murphy}, Tara and {Lenc}, Emil and {Walker}, Mark and {Bannister}, Keith and {Kaplan}, David L. and {Mahony}, Elizabeth K.},
        title = "{ASKAP observations of multiple rapid scintillators reveal a degrees-long plasma filament}",
      journal = {\mnras},
         year = 2021,
        month = apr,
       volume = {502},
       number = {3},
        pages = {3294-3311},
          doi = {10.1093/mnras/stab139},
archivePrefix = {arXiv},
       eprint = {2101.06048},
 primaryClass = {astro-ph.GA},
       adsurl = {https://ui.adsabs.harvard.edu/abs/2021MNRAS.502.3294W}
}

@ARTICLE{2023MNRAS.523.5661W,
	author = {{Wang}, Yuanming and {Murphy}, Tara and {Lenc}, Emil and {Mercorelli}, Louis and {Driessen}, Laura and {Pritchard}, Joshua and {Lao}, Baoqiang and {Kaplan}, David L. and {An}, Tao and {Bannister}, Keith W. and {Heald}, George and {Lu}, Shuoying and {Tuntsov}, Artem and {Walker}, Mark and {Zic}, Andrew},
	title = "{Radio variable and transient sources on minute time-scales in the ASKAP pilot surveys}",
	journal = {\mnras},
	year = 2023,
	month = aug,
	volume = {523},
	number = {4},
	pages = {5661-5680},
	doi = {10.1093/mnras/stad1727},
	archivePrefix = {arXiv},
	eprint = {2306.04263},
	primaryClass = {astro-ph.HE},
	adsurl = {https://ui.adsabs.harvard.edu/abs/2023MNRAS.523.5661W}
}

@ARTICLE{2020A&A...641L...4O,
       author = {{Oosterloo}, T.~A. and {Vedantham}, H.~K. and {Kutkin}, A.~M. and {Adams}, E.~A.~K. and {Adebahr}, B. and {Coolen}, A.~H.~W.~M. and {Damstra}, S. and {de Blok}, W.~J.~G. and {D{\'e}nes}, H. and {Hess}, K.~M. and {Hut}, B. and {Loose}, G.~M. and {Lucero}, D.~M. and {Maan}, Y. and {Morganti}, R. and {Moss}, V.~A. and {Mulder}, H. and {Norden}, M.~J. and {Offringa}, A.~R. and {Oostrum}, L.~C. and {Orr{\`u}}, E. and {Ruiter}, M. and {Schulz}, R. and {van den Brink}, R.~H. and {van der Hulst}, J.~M. and {van Leeuwen}, J. and {Vermaas}, N.~J. and {Vohl}, D. and {Wijnholds}, S.~J. and {Ziemke}, J.},
        title = "{Extreme intra-hour variability of the radio source J1402+5347 discovered with Apertif}",
      journal = {\aap},
         year = 2020,
        month = sep,
       volume = {641},
          eid = {L4},
        pages = {L4},
          doi = {10.1051/0004-6361/202038378},
archivePrefix = {arXiv},
       eprint = {2008.07945},
 primaryClass = {astro-ph.GA},
       adsurl = {https://ui.adsabs.harvard.edu/abs/2020A&A...641L...4O}
}

@ARTICLE{2019ARA&A..57..467B,
       author = {{Blandford}, Roger and {Meier}, David and {Readhead}, Anthony},
        title = "{Relativistic Jets from Active Galactic Nuclei}",
      journal = {\araa},
         year = 2019,
        month = aug,
       volume = {57},
        pages = {467-509},
          doi = {10.1146/annurev-astro-081817-051948},
archivePrefix = {arXiv},
       eprint = {1812.06025},
 primaryClass = {astro-ph.HE},
       adsurl = {https://ui.adsabs.harvard.edu/abs/2019ARA&A..57..467B}
}

@ARTICLE{2016Sci...351..354B,
       author = {{Bannister}, Keith W. and {Stevens}, Jamie and {Tuntsov}, Artem V. and {Walker}, Mark A. and {Johnston}, Simon and {Reynolds}, Cormac and {Bignall}, Hayley},
        title = "{Real-time detection of an extreme scattering event: Constraints on Galactic plasma lenses}",
      journal = {Science},
         year = 2016,
        month = jan,
       volume = {351},
       number = {6271},
        pages = {354-356},
          doi = {10.1126/science.aac7673},
archivePrefix = {arXiv},
       eprint = {1601.05876},
 primaryClass = {astro-ph.GA},
       adsurl = {https://ui.adsabs.harvard.edu/abs/2016Sci...351..354B}
}

@ARTICLE{2001MNRAS.325.1411K,
       author = {{Kedziora-Chudczer}, L. and {Jauncey}, D.~L. and {Wieringa}, M.~H. and {Tzioumis}, A.~K. and {Reynolds}, J.~E.},
        title = "{The ATCA intraday variability survey of extragalactic radio sources}",
      journal = {\mnras},
         year = 2001,
        month = aug,
       volume = {325},
       number = {4},
        pages = {1411-1430},
          doi = {10.1046/j.1365-8711.2001.04516.x},
archivePrefix = {arXiv},
       eprint = {astro-ph/0103506},
 primaryClass = {astro-ph},
       adsurl = {https://ui.adsabs.harvard.edu/abs/2001MNRAS.325.1411K}
}

@ARTICLE{1997ApJ...490L...9K,
       author = {{Kedziora-Chudczer}, L. and {Jauncey}, D.~L. and {Wieringa}, M.~H. and {Walker}, M.~A. and {Nicolson}, G.~D. and {Reynolds}, J.~E. and {Tzioumis}, A.~K.},
        title = "{PKS 0405-385: The Smallest Radio Quasar?}",
      journal = {\apjl},
         year = 1997,
        month = nov,
       volume = {490},
       number = {1},
        pages = {L9-L12},
          doi = {10.1086/311001},
archivePrefix = {arXiv},
       eprint = {astro-ph/9710057},
 primaryClass = {astro-ph},
       adsurl = {https://ui.adsabs.harvard.edu/abs/1997ApJ...490L...9K}
}

@ARTICLE{1987AJ.....94.1493H,
       author = {{Heeschen}, D.~S. and {Krichbaum}, Th. and {Schalinski}, C.~J. and {Witzel}, A.},
        title = "{Rapid Variability of Extragalactic Radio Sources}",
      journal = {\aj},
         year = 1987,
        month = dec,
       volume = {94},
        pages = {1493},
          doi = {10.1086/114583},
       adsurl = {https://ui.adsabs.harvard.edu/abs/1987AJ.....94.1493H}
}

@ARTICLE{2008ApJ...689..108L,
       author = {{Lovell}, J.~E.~J. and {Rickett}, B.~J. and {Macquart}, J.-P. and {Jauncey}, D.~L. and {Bignall}, H.~E. and {Kedziora-Chudczer}, L. and {Ojha}, R. and {Pursimo}, T. and {Dutka}, M. and {Senkbeil}, C. and {Shabala}, S.},
        title = "{The Micro-Arcsecond Scintillation-Induced Variability (MASIV) Survey. II. The First Four Epochs}",
      journal = {\apj},
         year = 2008,
        month = dec,
       volume = {689},
       number = {1},
        pages = {108-126},
          doi = {10.1086/592485},
archivePrefix = {arXiv},
       eprint = {0808.1140},
 primaryClass = {astro-ph},
       adsurl = {https://ui.adsabs.harvard.edu/abs/2008ApJ...689..108L}
}

@ARTICLE{1981ApJ...246...91D,
       author = {{Dennison}, B. and {Condon}, J.~J.},
        title = "{A search for interstellar scintillations in a large sample of low-frequency variable sources.}",
      journal = {\apj},
         year = 1981,
        month = may,
       volume = {246},
        pages = {91-99},
          doi = {10.1086/158901},
       adsurl = {https://ui.adsabs.harvard.edu/abs/1981ApJ...246...91D}
}

@ARTICLE{2014AJ....147...14D,
       author = {{Deller}, A.~T. and {Middelberg}, E.},
        title = "{mJIVE-20: A Survey for Compact mJy Radio Objects with the Very Long Baseline Array}",
      journal = {\aj},
         year = 2014,
        month = jan,
       volume = {147},
       number = {1},
          eid = {14},
        pages = {14},
          doi = {10.1088/0004-6256/147/1/14},
archivePrefix = {arXiv},
       eprint = {1310.8191},
 primaryClass = {astro-ph.CO},
       adsurl = {https://ui.adsabs.harvard.edu/abs/2014AJ....147...14D}
}

@article{10.1111/j.1365-2966.2004.08159.x,
    author = {Walker, M. A. and Melrose, D. B. and Stinebring, D. R. and Zhang, C. M.},
    title = {Interpretation of parabolic arcs in pulsar secondary spectra},
    journal = {Monthly Notices of the Royal Astronomical Society},
    volume = {354},
    number = {1},
    pages = {43-54},
    year = {2004},
    month = {10},
    issn = {0035-8711},
    doi = {10.1111/j.1365-2966.2004.08159.x},
    url = {https://doi.org/10.1111/j.1365-2966.2004.08159.x},
    eprint = {https://academic.oup.com/mnras/article-pdf/354/1/43/2862071/354-1-43.pdf},
}

@ARTICLE{2019NatAs...3..387P,
       author = {{Panessa}, Francesca and {Baldi}, Ranieri Diego and {Laor}, Ari and {Padovani}, Paolo and {Behar}, Ehud and {McHardy}, Ian},
        title = "{The origin of radio emission from radio-quiet active galactic nuclei}",
      journal = {Nature Astronomy},
         year = 2019,
        month = apr,
       volume = {3},
        pages = {387-396},
          doi = {10.1038/s41550-019-0765-4},
archivePrefix = {arXiv},
       eprint = {1902.05917},
 primaryClass = {astro-ph.GA},
       adsurl = {https://ui.adsabs.harvard.edu/abs/2019NatAs...3..387P}
}

@ARTICLE{2025A&A...698A.157O,
       author = {{Orienti}, M. and {D'Ammando}, F. and {Dallacasa}, D. and {Migliori}, G. and {Rossi}, P. and {Bodo}, G.},
        title = "{VLBA observations of a sample of low-power compact symmetric objects}",
      journal = {\aap},
         year = 2025,
        month = jun,
       volume = {698},
          eid = {A157},
        pages = {A157},
          doi = {10.1051/0004-6361/202553798},
archivePrefix = {arXiv},
       eprint = {2505.03461},
 primaryClass = {astro-ph.GA},
       adsurl = {https://ui.adsabs.harvard.edu/abs/2025A&A...698A.157O}
}

@ARTICLE{1987Natur.326..675F,
       author = {{Fiedler}, R.~L. and {Dennison}, B. and {Johnston}, K.~J. and {Hewish}, A.},
        title = "{Extreme scattering events caused by compact structures in the interstellar medium}",
      journal = {\nat},
         year = 1987,
        month = apr,
       volume = {326},
       number = {6114},
        pages = {675-678},
          doi = {10.1038/326675a0},
       adsurl = {https://ui.adsabs.harvard.edu/abs/1987Natur.326..675F}
}

@ARTICLE{2016A&ARv..24...13P,
       author = {{Padovani}, Paolo},
        title = "{The faint radio sky: radio astronomy becomes mainstream}",
      journal = {\aapr},
         year = 2016,
        month = sep,
       volume = {24},
       number = {1},
          eid = {13},
        pages = {13},
          doi = {10.1007/s00159-016-0098-6},
archivePrefix = {arXiv},
       eprint = {1609.00499},
 primaryClass = {astro-ph.GA},
       adsurl = {https://ui.adsabs.harvard.edu/abs/2016A&ARv..24...13P}
}

@ARTICLE{2016ApJ...817..176T,
       author = {{Tuntsov}, Artem V. and {Walker}, Mark A. and {Koopmans}, Leon V.~E. and {Bannister}, Keith W. and {Stevens}, Jamie and {Johnston}, Simon and {Reynolds}, Cormac and {Bignall}, Hayley E.},
        title = "{Dynamic Spectral Mapping of Interstellar Plasma Lenses}",
      journal = {\apj},
         year = 2016,
        month = feb,
       volume = {817},
       number = {2},
          eid = {176},
        pages = {176},
          doi = {10.3847/0004-637X/817/2/176},
archivePrefix = {arXiv},
       eprint = {1512.03411},
 primaryClass = {astro-ph.IM},
       adsurl = {https://ui.adsabs.harvard.edu/abs/2016ApJ...817..176T}
}

@ARTICLE{2017MNRAS.469.5023T,
       author = {{Tuntsov}, Artem V. and {Stevens}, Jamie and {Bannister}, Keith W. and {Bignall}, Hayley and {Johnston}, Simon and {Reynolds}, Cormac and {Walker}, Mark A.},
        title = "{Scintillation kinks, bumps and wiggles in the radio spectrum of the quasar PMN J1106-3647}",
      journal = {\mnras},
         year = 2017,
        month = aug,
       volume = {469},
       number = {4},
        pages = {5023-5032},
          doi = {10.1093/mnras/stx1223},
archivePrefix = {arXiv},
       eprint = {1705.06051},
 primaryClass = {astro-ph.HE},
       adsurl = {https://ui.adsabs.harvard.edu/abs/2017MNRAS.469.5023T}
}

@ARTICLE{2006A&A...446..185M,
       author = {{Macquart}, J.-P. and {de Bruyn}, A.~G.},
        title = "{Diffractive interstellar scintillation of the quasar J1819+3845 at {\ensuremath{\lambda}}21 cm}",
      journal = {\aap},
         year = 2006,
        month = jan,
       volume = {446},
       number = {1},
        pages = {185-200},
          doi = {10.1051/0004-6361:20053293},
archivePrefix = {arXiv},
       eprint = {astro-ph/0510495},
 primaryClass = {astro-ph},
       adsurl = {https://ui.adsabs.harvard.edu/abs/2006A&A...446..185M}
}

@ARTICLE{2021MNRAS.501.6139R,
       author = {{Ross}, K. and {Callingham}, J.~R. and {Hurley-Walker}, N. and {Seymour}, N. and {Hancock}, P. and {Franzen}, T.~M.~O. and {Morgan}, J. and {White}, S.~V. and {Bell}, M.~E. and {Patil}, P.},
        title = "{Spectral variability of radio sources at low frequencies}",
      journal = {\mnras},
         year = 2021,
        month = mar,
       volume = {501},
       number = {4},
        pages = {6139-6155},
          doi = {10.1093/mnras/staa3795},
archivePrefix = {arXiv},
       eprint = {2012.01842},
 primaryClass = {astro-ph.GA},
       adsurl = {https://ui.adsabs.harvard.edu/abs/2021MNRAS.501.6139R}
}

@ARTICLE{2023MNRAS.523.2219A,
       author = {{Andersson}, Alex and {Lintott}, Chris and {Fender}, Rob and {Bright}, Joe and {Carotenuto}, Francesco and {Driessen}, Laura and {Espinasse}, Mathilde and {Gasealahwe}, Kelebogile and {Heywood}, Ian and {van der Horst}, Alexander J. and et al.},
        title = "{Bursts from Space: MeerKAT - the first citizen science project dedicated to commensal radio transients}",
      journal = {\mnras},
         year = 2023,
        month = aug,
       volume = {523},
       number = {2},
        pages = {2219-2235},
          doi = {10.1093/mnras/stad1298},
archivePrefix = {arXiv},
       eprint = {2304.14157},
 primaryClass = {astro-ph.HE},
       adsurl = {https://ui.adsabs.harvard.edu/abs/2023MNRAS.523.2219A}
}

@misc{murphy2025dawesreview13new,
      title={The Dawes Review 13: A New Look at The Dynamic Radio Sky}, 
      author={Tara Murphy and David L. Kaplan},
      year={2025},
      eprint={2511.10785},
      archivePrefix={arXiv},
      primaryClass={astro-ph.SR},
      url={https://arxiv.org/abs/2511.10785}, 
}

@ARTICLE{2017ApJ...845...89V,
       author = {{Vedantham}, H.~K. and {Readhead}, A.~C.~S. and {Hovatta}, T. and {Pearson}, T.~J. and {Blandford}, R.~D. and {Gurwell}, M.~A. and {L{\"a}hteenm{\"a}ki}, A. and {Max-Moerbeck}, W. and {Pavlidou}, V. and {Ravi}, V. and {Reeves}, R.~A. and {Richards}, J.~L. and {Tornikoski}, M. and {Zensus}, J.~A.},
        title = "{Symmetric Achromatic Variability in Active Galaxies: A Powerful New Gravitational Lensing Probe?}",
      journal = {\apj},
         year = 2017,
        month = aug,
       volume = {845},
       number = {2},
          eid = {89},
        pages = {89},
          doi = {10.3847/1538-4357/aa745c},
archivePrefix = {arXiv},
       eprint = {1702.06582},
 primaryClass = {astro-ph.HE},
       adsurl = {https://ui.adsabs.harvard.edu/abs/2017ApJ...845...89V}
}

@article{Kiehlmann_2024,
doi = {10.3847/1538-4357/ad0c56},
url = {https://doi.org/10.3847/1538-4357/ad0c56},
year = {2024},
month = {jan},
publisher = {The American Astronomical Society},
volume = {961},
number = {2},
pages = {240},
author = {Kiehlmann, S. and Lister, M. L. and Readhead, A. C. S and Liodakis, I. and O’Neill, Sandra and Pearson, T. J. and Sheldahl, Evan and Siemiginowska, Aneta and Tassis, K. and Taylor, G. B. and Wilkinson, P. N.},
title = {Compact Symmetric Objects. I. Toward a Comprehensive Bona Fide Catalog},
journal = {The Astrophysical Journal}
}

@ARTICLE{2024MNRAS.528.6292J,
       author = {{Jow}, Dylan L. and {Pen}, Ue-Li and {Baker}, Daniel},
        title = "{On the cusp of cusps: a universal model for extreme scattering events in the ISM}",
      journal = {\mnras},
         year = 2024,
        month = mar,
       volume = {528},
       number = {4},
        pages = {6292-6301},
          doi = {10.1093/mnras/stae300},
archivePrefix = {arXiv},
       eprint = {2301.08344},
 primaryClass = {astro-ph.HE},
       adsurl = {https://ui.adsabs.harvard.edu/abs/2024MNRAS.528.6292J}
}

@ARTICLE{2025ApJ...982...61S,
       author = {{Suvorov}, Arthur G. and {Walker}, Mark A.},
        title = "{Tidal Disruption of ``Snow Clouds'' by Unassociated Stars}",
      journal = {\apj},
         year = 2025,
        month = mar,
       volume = {982},
       number = {1},
          eid = {61},
        pages = {61},
          doi = {10.3847/1538-4357/adb74b},
archivePrefix = {arXiv},
       eprint = {2501.17419},
 primaryClass = {astro-ph.GA},
       adsurl = {https://ui.adsabs.harvard.edu/abs/2025ApJ...982...61S}
}

@incollection{TaoAn03.2026.SKA, author = {Tao An and author2 and author3 and author4 and author5},title = {},year = {2026},publisher = {},note = {arXiv search: Report number AASKAII/TaoAn03},booktitle = {Advancing Astrophysics with the SKA -- II (AASKAII)}}

@incollection{Shu01.2026.SKA, author = {Xinwen Shu and author2 and author3 and author4 and author5},title = {},year = {2026},publisher = {},note = {arXiv search: Report number AASKAII/Shu01},booktitle = {Advancing Astrophysics with the SKA -- II (AASKAII)}}

@incollection{Castignani01.2026.SKA, author = {Gianluca Castignani and author2 and author3 and author4 and author5},title = {},year = {2026},publisher = {},note = {arXiv search: Report number AASKAII/Castignani01},booktitle = {Advancing Astrophysics with the SKA -- II (AASKAII)}}

@incollection{Rosch01.2026.SKA, author = {Florian Rösch and author2 and author3 and author4 and author5},title = {},year = {2026},publisher = {},note = {arXiv search: Report number AASKAII/Rosch01},booktitle = {Advancing Astrophysics with the SKA -- II (AASKAII)}}

@incollection{Chhetri01.2026.SKA, author = {Rajan Chhetri and author2 and author3 and author4 and author5},title = {},year = {2026},publisher = {},note = {arXiv search: Report number AASKAII/Chhetri01},booktitle = {Advancing Astrophysics with the SKA -- II (AASKAII)}}

@incollection{AlexAndersson01.2026.SKA, author = {Alex Andersson and author2 and author3 and author4 and author5},title = {},year = {2026},publisher = {},note = {arXiv search: Report number AASKAII/AlexAndersson01},booktitle = {Advancing Astrophysics with the SKA -- II (AASKAII)}}

@incollection{Kadler01.2026.SKA, author = {Matthias Kadler and author2 and author3 and author4 and author5},title = {},year = {2026},publisher = {},note = {arXiv search: Report number AASKAII/Kadler01},booktitle = {Advancing Astrophysics with the SKA -- II (AASKAII)}}

@incollection{Tiburzi01.2026.SKA, author = {Caterina Tiburzi and author2 and author3 and author4 and author5},title = {},year = {2026},publisher = {},note = {arXiv search: Report number AASKAII/Tiburzi01},booktitle = {Advancing Astrophysics with the SKA -- II (AASKAII)}}

\end{document}